\documentclass[aps,prl,twocolumn,groupedaddress,showpacs]{revtex4}
\usepackage{amsmath}
\usepackage{amsfonts}
\usepackage{amssymb}
\usepackage{graphicx}
\usepackage{verbatim}
\usepackage{xcolor}
\usepackage[normalem]{ulem}
\newcommand{\ts}{\textsuperscript}
\usepackage{epstopdf}

\begin{document}

\title{Generation of high harmonics from silicon}

\author{G. Vampa\ts{1}}
\email[]{gvamp015@uottawa.ca}
\author{T. J. Hammond\ts{1}}
\author{N. Thir\'{e}\ts{2}}
\author{B. E. Schmidt\ts{2}}
\author{F. L\'{e}gar\'{e}\ts{2}}
\author{D. D. Klug\ts{3}}
\author{P. B. Corkum\ts{1,3}}

\affiliation{\ts{1}Department of Physics, University of Ottawa, Ottawa, ON K1N 6N5, Canada}
\affiliation{\ts{2}INRS-EMT, 1650 Boulevard Lionel-Boulet, CP 1020, Varennes, Qu\'{e}bec J3X 1S2, Canada}
\affiliation{\ts{3}National Research Council of Canada, Ottawa, Ontario, Canada K1A 0R6}

\date{\today}


\begin{abstract}
\noindent
We generate high-order harmonics of a mid-infrared laser from a silicon single crystal and find their origin in the recollision of coherently accelerated electrons with their holes, analogously to the atomic and molecular case, and to ZnO [Vampa, {\it et al.}, Nature {\bf 522}, 462-464 (2015)], a direct bandgap material. Therefore indirect bandgap materials are shown to sustain the recollision process as well as direct bandgap materials. Furthermore, we find that the generation is perturbed with electric fields as low as 30 V/$\mu$m, equal to the DC damage threshold. Our results extend high-harmonic spectroscopy to the most technologically relevant material, and open the possibility to integrate high harmonics with conventional electronics.
\end{abstract}

\maketitle
\noindent
50 years ago Keldysh predicted that strong laser fields can adiabatically tunnel electrons in isolated atoms from a bound state to the vacuum \cite{keldysh}.  Tunnelling is followed by acceleration of the ionized electrons and their recollision with the atomic core, a process that emits high order harmonics of the driving laser field \cite{hhg}. 

Keldysh predicted that electrons in solids also undergo adiabatic tunnelling from the valence to the conduction band. Although tunnelling from atoms has been the matter of intense research ever since, less attention has been paid to solids. Recently, however, the step-like increase in conduction band population arising from tunnelling near each crest of the laser field has been measured in silicon, a material with a small indirect bandgap \cite{schultze}. In parallel with atoms, the generation of high harmonics from recolliding electron-hole pairs has also been demonstrated, but only from ZnO \cite{vampanature,ghimire}, a semiconductor with a wide direct bandgap, and with a mid-infrared laser. Following the demonstration of tunnelling from silicon with an 800nm laser \cite{schultze}, and because the adiabatic character of tunnelling increases at longer wavelengths, recollision-based high harmonics employing a mid-infrared laser should be generated from silicon as well.


If harmonics from silicon are measured, however, it is important to establish their origin. Are they emitted from a single band \cite{blochosc}, or from recombination of the electron-hole pair between two bands \cite{vampaprl}, or even from the interplay between multiple bands \cite{hubernature,gaarde,hawkins}?\\

Here we present the generation of high harmonics from a silicon single crystal, that extend from the mid-infrared to the ultra-violet spectral region. We characterize the generation process and find that:\\
(i) recolliding electron-hole pairs dominate the emission, thus extending the importance of recollisions to indirect bandgap materials;\\
(ii) the generation is sensitive to perturbing fields as low as 30 V/$\mu$m, equal to the DC breakdown field\cite{ioffre};\\
(iii) the generation and its sensitivity to the perturbing field strongly depend on the crystallographic orientation with respect to the linear laser polarization. Because silicon is isotropic, this dependence has a microscopic origin.\\


The generation mechanism is characterized by perturbing it with a second harmonic field, thereby producing even harmonics \cite{nirit,lund}. Their strength modulates as the second harmonic is delayed relative to the fundamental. The optimum phase that maximizes the intensity of the even harmonics differs between harmonic orders. In a recollision process, different harmonic orders map to trajectories of the electron and the hole that differ in their times of creation and recollision. The variation of the optimum phase with harmonic order encodes the different trajectories, and therefore identifies recollision of the electron with the hole as the dominant emission channel \cite{vampanature}. \\

\begin{figure}
\includegraphics[scale=1.0]{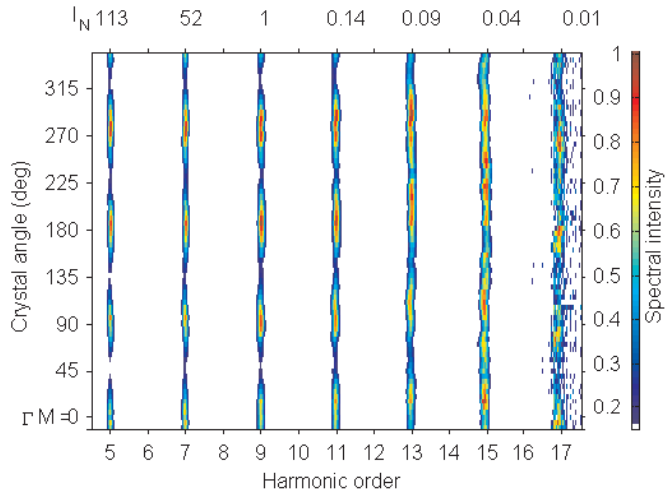}
\caption{\label{figure1}High harmonic spectra modulate as the linear polarization is rotated, with maximum emission for polarization parallel to the [110] direction. Each harmonic order is normalized to 1 for clarity. Their original intensity as compared to the 9th order is indicated at the top.}
\end{figure}


In a first experiment we focus mid-infrared linearly polarized laser pulses of 100 fs duration with a central wavelength of 3.5 $\mu$m (0.35 eV photon energy) in a 500 nm thick film of single crystal silicon, at an intensity of 0.10 V/\AA in the crystal. The film is epitaxially grown on a 500 $\mu$m thick R-plane Sapphire substrate. The beam is incident on the silicon side. The [100] orientation of the Si lattice is parallel to the projection of the c-axis of Sapphire on the surface \cite{siepitaxy}. Measurement of the birefringence of the substrate determines the orientation of this projection with respect to the laser polarization, which ultimately determines the orientation of the Si lattice with respect to the latter. The field strength in the material is calculated from the vacuum field strength accounting for reflection losses at the air-silicon interface: $F_{si} = 2F_{vac}/(n_{Si}+1)$. Here, $n_{si}=3.4286$. The surface normal is aligned to the [001] direction, and is parallel to the laser k-vector. 

\begin{figure}
\includegraphics[scale=1]{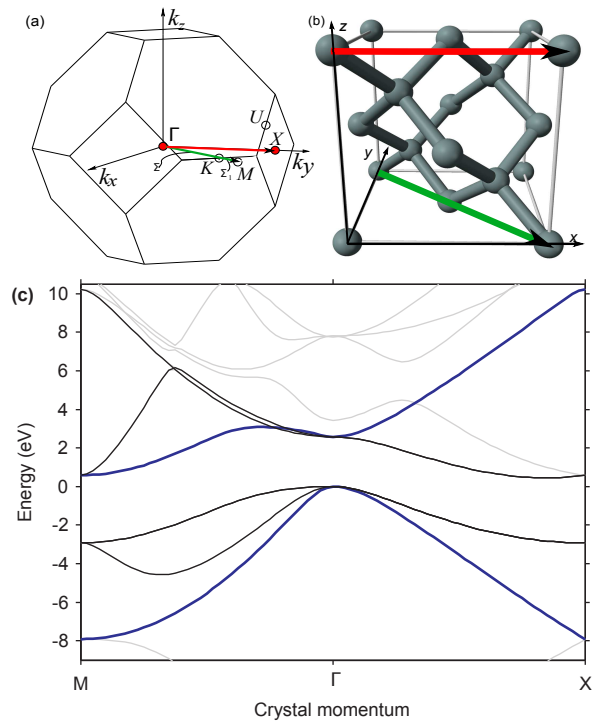}
\caption{\label{figure2}(a) First Brillouin Zone of the FCC silicon lattice; The $M$ point lies outside the first Brillouin Zone, along the $\Gamma-K$ line \cite{bilbao1,bilbao2}. (b) The lattice of silicon. The laser propagates along the z-axis. When the laser is polarized along the [100] direction (red arrow in (b)), electrons are accelerated along the $\Gamma-X$ direction of the Brillouin Zone (red arrow in (a)), and when it is polarized along the [110] (green arrow in (b)) direction they move along the $\Gamma-M$ direction (green arrow in (a)). (c) Band structure of silicon obtained from {\it ab initio} calculations. Bands coloured in blue have been considered for the classical calculation of Fig. \ref{figure4}. Bands degenerate with the blue one at $\Gamma$ are coloured in black.}
\end{figure}

Before we start, we introduce the dependence of high-harmonic spectra in the presence of the fundamental field alone on the crystallographic orientation with respect to the linear laser polarization.
When the linear polarization is rotated about the [001] axis of the crystal, high-harmonic spectra modulate with 4-fold symmetry, as shown in Fig. \ref{figure1}. All harmonics maximize when the polarization of the fundamental beam is oriented along the [110] direction. For this orientation, electron-hole pairs created by the strong laser field are accelerated along the projection of the Si-Si bonds onto the (001) plane, as indicated by the green arrow in Fig. \ref{figure2}b, or along the $\Gamma-K-M$ direction in the reciprocal space (green arrow in  Fig. \ref{figure2}a). The $M$ point is equivalent to the $X$ point, but it is reached along the $\Sigma_1$ line. High harmonics are suppressed for polarization along the [100] direction, when electron-hole pairs move in and out of the molecular chain, as indicated by the red arrow in Fig. \ref{figure2}b. In reciprocal space, the polarization is along the $\Gamma-X$ direction (red arrow in Fig. \ref{figure2}a).\\

Because the linear optical properties of silicon are isotropic, neither absorption nor phase-matching explains the strong rotational dependence of the harmonics. The observed behaviour can only originate from the microscopic generation process.\\

\begin{figure}
\includegraphics[scale=1]{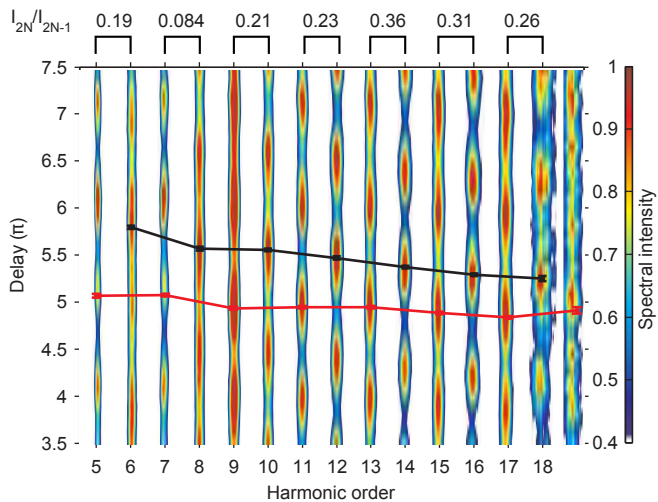}
\caption{\label{figure3}The harmonic intensity modulates when the delay between the fundamental and the second harmonic is varied. While the phase of the oscillation for the odd harmonics (red line) is almost constant, the phase of the even harmonics (black line) varies with harmonic order, for polarization along the [110] direction. The relative intensity of even and odd harmonics is shown at the top. Each harmonic order is separately normalized.}
\end{figure}

We can gain additional insight into the high-harmonic generation mechanism by perturbing the process with a second harmonic field \cite{nirit,vampanature}. When the fundamental and its second harmonic are properly phased, even harmonics are produced, as shown in Fig. \ref{figure3} for a second harmonic intensity of $9 \times 10^{-4}$ of the fundamental (corresponding to a field strength of only 30 V/$\mu$m). In this case, laser pulses of 55 fs duration with a central wavelength of 3.7 $\mu$m (0.33 eV photon energy) are focused in a 40 nm thin film of single crystal silicon at the same intensity as in Fig. \ref{figure1}. The film is plasma etched from its initial thickness of 500 nm. In this experiment, the beam is incident on the Sapphire side, but to avoid birefringence the polarization is set parallel to the [110] direction.\\

When the phase between the two colours is varied, the even harmonic intensity modulates. The resulting spectrogram is shown in Fig. \ref{figure3} for polarization along the $\Gamma-M$ direction. The intensity of the even harmonics relative to the odd ones is reported at the top of the graph. The visibility of the modulation, defined as $V = 2 \times (I_{max}-I_{min})/(I_{max}+I_{min})$, steadily increases with harmonic order from 0.04 to 0.8. The phase of the modulation of the even harmonics differs between harmonic orders (the black line connects the maxima of the even harmonics). A similar behaviour is observed in atomic and molecular high-harmonic generation \cite{nirit,lund}, and in ZnO crystals \cite{vampanature}. This signature confirms that recolliding electron-hole pairs and their recombination are the dominant source of harmonics in silicon for excitation with mid-infrared lasers.\\

\begin{figure}
\includegraphics[scale=1]{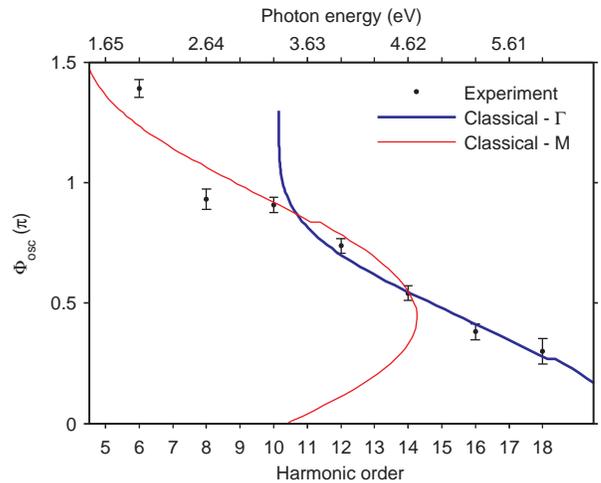}
\caption{\label{figure4}Optimum phase for electron-hole pairs tunnelling at $\Gamma$ (blue line), for polarization along the $\Gamma-M$ direction (as in Fig. \ref{figure3}). The minimum harmonic energy corresponds to the direct bandgap of silicon (10$^{th}$ harmonic order). The classical calculation doesn't allow recombination below the minimum bandgap. The red line shows the optimum phase for electrons that tunnel at $M$, instead. Here, the lowest band gap is close to the minimum (indirect) band gap of silicon (1.1 eV).}
\end{figure}

To quantitatively assess the experimental results in Fig. \ref{figure4}, we use the analytical solution of the recollision dipole \cite{vampaprl}. It suggests that a second harmonic field adds the following phase to the dipole \cite{vampanature}:
\begin{equation}
\label{phase}
\sigma(t,\phi)=\int_{t'}^t \Delta\mathbf{v}(\tau,t')\mathbf{A_2}(\tau,\phi)d\tau
\end{equation}
where $t'$ and $t$ are the birth and recollision times (specific to each harmonic order), $\Delta\mathbf{v}=\mathbf{v_e}-\mathbf{v_h}=\mathbf{\nabla_k}E_g(\mathbf{k})$ is the difference in the band velocities between electrons and holes in the fundamental field alone (with $E_g = E_c - E_v$ the momentum-dependent bandgap), $\mathbf{A_2}$ is the second harmonic field amplitude and $\phi$ the phase difference with the fundamental field. The crystal momentum is ${\bf k} = {\bf A_1}(\tau) - {\bf A_1}(t')$, where ${\bf A_1}$ is the vector potential of the fundamental field. The even harmonic intensity modulates as $\sin^2(\sigma)$ \cite{dahlstrom}. Following the procedure outlined in Ref. \cite{nirit}, $\sigma$  can be decomposed as $\sigma(t,\phi)=\sigma_s(t)\cos(\phi) + \sigma_c(t)\sin(\phi)$, and the phase of the modulation calculated as $\Phi_{osc}=\arctan(\sigma_c/\sigma_s)$. When all electron-hole pair trajectories launched along the $\Gamma-M$ direction between the peak and the following zero of the fundamental field are calculated (each trajectory leading to emission at different harmonic photon energies), the phase of the modulation plotted in blue in Fig. \ref{figure4} is obtained. To calculate the electron's and hole's velocities, required by Eq. (\ref{phase}), we assume that tunnelling only happens at $\Gamma$ between the bands with the smallest effective mass. These bands are coloured blue in Fig. \ref{figure2}c.\\

There are two important comments.\\
First, as in atomic media experiments, in this experiment the delay between the fundamental and the second harmonic has not been measured with sub-cycle precision (this measurement is possible in principle by replacing silicon with a thin nonlinear crystal). Therefore, the experimental data is known up to a constant phase. In Fig. \ref{figure4} the experimental absolute phase has been taken equal to that calculated at the 14$^{th}$ harmonic order.\\
Second, the classical calculation does not allow recombination to occur with energy smaller than the minimum (direct) bandgap. Therefore, the calculated $\Phi_{osc}$ does not extend below the 10$^{th}$ harmonic order. We see the analogous limit in the semiclassical analysis of gases.\\
Within these limits, the agreement between experiment and analysis is excellent.\\

The model of high harmonic generation presented so far assumes tunnelling at the direct band gap. In an alternative scenario, electrons pick-up crystal momentum from phonons during tunnelling and start their motion from the minimum energy state away from $\Gamma$ in the conduction band. Then, electrons and holes accelerate in reciprocal space by the same amount, but remain separated by the initial momentum offset. In real space electrons and holes are initially at rest because electrons tunnel from the valence band maximum to a local minimum of the conduction band, where ${\rm v_{c,v}}=\nabla_{\bf \rm k} E_{\rm {c,v}}({\bf\rm k}) = 0$. Then they begin to accelerate away from each other. Following their re-encounter later in the laser cycle, emission of a high harmonic photon requires recombination with exchange of the same phonons absorbed during tunnelling. This scenario and the one in which tunnelling happens at $\Gamma$ can be easily discriminated because electrons travel different portions of the Brillouin zone at different times. Therefore, $\Phi_{osc}$ is different. The red line of Fig. \ref{figure4} shows $\Phi_{osc}$ for electrons that tunnel at $M$ in the conduction band. Because the minimum gap is now close to the indirect band gap of silicon (1.1 eV), the classical calculation extends below the 10$^{th}$ harmonic. The cut off extends only up to the 14$^{th}$ order, contrary to the experiment. Therefore, our analysis confirms that tunnelling only sees direct band gaps \cite{schultze}.\\

In conclusion, our experiment shows that strong field tunnelling in silicon is followed by emission of high harmonics as a result of the acceleration of electrons and holes and their subsequent recollision and recombination.\\

Our results are technologically relevant. Like for ZnO, we find that harmonics from silicon are sensitive to very weak electric fields (as low as 30 V/$\mu$m, equal to the DC breakdown threshold), and comparable to those found in electronic circuits. It is possible to break the symmetry, thereby creating even harmonics, with DC fields. We could exploit the unique use of silicon for electronic components and integrated photonics \cite{rickman} to record live images of working circuits, an alternative to conventional probing by second harmonic generation \cite{currentind}. In our method, odd harmonics would be generated by the circuit at rest. A propagating electric field would break the symmetry and generate even harmonics, which would diffract according to the instantaneous distribution of the fields within the circuit. Propagation of optical pulses in silicon photonics circuits could also be probed. The method benefits from high spatial resolution, dictated by the wavelength of the diffracted harmonics, and sub-laser-cycle temporal resolution.

Finally, in the brief analysis of Fig. \ref{figure4} we assumed that only two bands contribute to the process. Given the complexity of the bandstructure near the $\Gamma$ point, it is likely that multiple bands contribute to the tunneling step \cite{hubernature}. When multiple molecular orbitals are ionized in gas phase experiments, attosecond electron and hole wavepackets are launched \cite{attosecondwp}. Silicon can therefore be a suitable platform for studying attosecond wavepacket dynamics in the condensed phase.\\

It is our pleasant duty to thank A. Laram\'{e}e, from the Advanced Laser Light Source, for his technical assistance during the experiment. The authors also acknowledge valuable support from The US AFOSR and Canada's NRC, NSERC and CFI. F. L\'{e}gar\'{e} acknowledges support from CFI-MSI.

\end{document}